\documentclass[english]{revtex4-1}
\usepackage[T1]{fontenc}
\usepackage[latin9]{inputenc}
\usepackage{color}
\usepackage{babel}
\usepackage{verbatim}
\usepackage{amsmath}
\usepackage{amssymb}
\usepackage{graphicx}
\usepackage[unicode=true,pdfusetitle,
 bookmarks=true,bookmarksnumbered=true,bookmarksopen=false,
 breaklinks=true,pdfborder={0 0 1},backref=false,colorlinks=true]
 {hyperref}
\hypersetup{
 linkcolor=red, citecolor=blue}
\begin{document}
\title{Spin Boltzmann equation for non-relativistic spin-1/2 fermions}
\author{Wen-Bo Dong, Yi-Liang Yin, Qun Wang}
\affiliation{Department of Modern Physics, University of Science and Technology
of China, Hefei, Anhui 230026, China}
\begin{abstract}
We derive the spin Boltzmann equations for spin-1/2 fermions in a
non-relativistic model with four-fermion contact interaction which
conserves spin degrees of freedom. A great advantage of the model
is that the spin matrix elements in collision terms can be completely
worked out and be put into such a compact form that one can clearly
see how spins are coupled in particle scatterings. A semi-classical
expansion in the Planck constant has been made and the on-shell part
of the spin Boltzmann equation up to the next-to-leading order is
derived. At the leading order the equilibrium spin distribution can
be obtained from the vanishing of the collision term for the spin
density. The spin chemical potential emerges as a natural consequence
of spin conservation. The off-shell part of the spin Boltzmann equation
is also discussed. The work can be extended to more sophisticated
interaction such as nuclear force in order to apply to spin polarization
phenomena in heavy-ion collisions at low energies.

\end{abstract}
\maketitle

\section{Introduction}

Very large orbital angular momenta (OAM) are generated in non-central
heavy-ion collisions which can be partially converted into the spin
polarization of hadrons along the direction of OAM or with respect
to the reaction plane \citep{Liang:2004ph,Liang:2004xn,Gao:2007bc}.
This effect is called the global spin polarization or global polarization
for short. The global polarization of $\Lambda$ hyperons (including
$\overline{\Lambda}$) has been measured for the first time by the
STAR collaboration in Au+Au collision at 200 GeV and lower energies
\citep{STAR:2017ckg,STAR:2018gyt}. The data show that the global
polarization is about 1.08$\pm$0.15\% ($\Lambda$) and 1.38$\pm$0.30\%
($\overline{\Lambda}$) with a decreasing behavior with the collision
energy.

Several theoretical methods have been developed for the global polarization.
These theoretical methods can be roughly put into three categories.
One category is related to the quantum statistical theory for particle
systems with spin degrees of freedom in equilibrium \citep{Becattini:2007sr,Becattini:2007nd,Becattini:2013fla,Becattini:2014yxa,Becattini:2018duy,Becattini:2019dxo}
{[}for a recent review, see, e.g., Ref. \citep{Becattini:2020sww}{]}.
One category is the microscopic transport theory based on kinetic
or Boltzmann equations for spin degrees of freedom \citep{Fang:2016vpj,Zhang:2019xya,Weickgenannt:2019dks,Gao:2019znl,Weickgenannt:2020aaf,Wang:2020pej,Sheng:2021kfc,Weickgenannt:2021cuo,Yang:2021fea,Sheng:2022ssd}
in terms of covariant Wigner functions \citep{Gao:2012ix,Chen:2012ca,Hidaka:2016yjf,Gao:2017gfq,Gao:2018wmr}
{[}see, e.g., Refs. \citep{Gao:2020pfu,Hidaka:2022dmn} for recent
reviews{]}. Another category is relativistic spin hydrodynamics \citep{Montenegro:2017rbu,Florkowski:2018fap,Montenegro:2018bcf,Becattini:2018duy,Hattori:2019lfp,Li:2020eon,Fukushima:2020ucl,Wang:2021ngp,Yi:2021unq,Weickgenannt:2022jes}
{[}see Ref. \citep{Florkowski:2018fap} for a review{]}, which incorporates
spin degrees of freedom into conventional relativistic hydrodynamics
applied to the strong interaction matter in heavy-ion collisions \citep{Kolb:2003dz,Heinz:2013th,Florkowski:2017olj,Romatschke:2017ejr}.
There are many phenomenological studies of the global and local polarization
using these theoretical methods to describe experimental data \citep{Karpenko:2016jyx,Li:2017slc,Xie:2017upb,Sun:2017xhx,Xia:2018tes,Wei:2018zfb,Baznat:2017jfj,Csernai:2018yok,Wu:2019eyi,Fu:2020oxj,Ivanov:2020qqe,Ivanov:2020wak,Ryu:2021lnx,Fu:2021pok,Becattini:2021iol,Yi:2021unq,Wu:2022mkr}
{[}for recent reviews, see, e.g., Refs. \citep{Gao:2020lxh,Huang:2020dtn,Becattini:2020ngo}{]}.

Recently HADES collaboration measured the global $\Lambda$ polarization
in Ag+Ag collisions at 2.55 GeV and Au+Au collisions at 2.4 GeV \citep{Kornas:2022cbl},
while STAR collaboration measured the same observable in Au+Au collisions
at 3 GeV \citep{PhysRevC.104.L061901}. Combining all these low energy
measurements with the high energy ones, the $\Lambda$ polarization
is observed to continue the increasing trend with decreasing collision
energy down to 2.4 GeV. At these collision energies of $\mathcal{O}(m_{N})$
where $m_{N}$ is the nucleon mass, the relativistic effect is small
and non-relativistic theory can be a proper approximation. Experiment
data can be described by models such as UrQMD and BUU \citep{PhysRevC.103.L031903,Deng:2021miw}.
These models are based on Boltzmann equations for hadrons which do
not incorporate spin degrees of freedom.

In this paper, we will derive the spin Boltzmann equation for spin-1/2
fermions in a non-relativistic model with four-fermion contact interaction
similar to the Nambu-Jona-Lasinio (NJL) model in relativistic theory
\citep{PhysRev.122.345,Nambu:1961fr}. The non-relativistic model
has a feature that the particle's spin is decoupled from its momentum
and is conserved in the interaction. This is very different from a
relativistic system in which the particle's spin and momentum are
entangled. The method is based on a previous work by one of us about
the relativistic system of spin-1/2 fermions \citep{Sheng:2021kfc}.
A great advantage of the current non-relativistic model is that the
spin matrix elements in collision terms can be completely worked out
and be put into such a compact form that one can clearly see how spins
are coupled in two-to-two scatterings of particles. This is not the
case in the relativistic theory \citep{Sheng:2021kfc}. The current
work can be extended to nucleon-nucleon interaction via nuclear force
and then can be applied to the global polarization in heavy-ion collisions
at low energies.

This paper is organized as follows. In Sect. \ref{sec:green-func},
we briefly introduce Green's functions in the CTP formalism. In Sect.
\ref{sec:Kadanoff-Baym-equation}, we derive the KB equation from
the Schwinger-Dyson equation in quasi-particle approximation. In Sect.
\ref{sec:NJL-on-shell}, we derive the on-shell part of the spin Boltzmann
equation at the leading and next-to-leading order in $\hbar$. In
Sect. \ref{sec:off-shell-part}, the off-shell part of the KB equation
is discussed. A summary of the results is given in the final section.


\section{Green functions for fermions in CTP formalism}

\label{sec:green-func}We consider a non-relativistic system of spin-1/2
fermions. A general form of the Lagrangian with four-fermion interaction
can be written as \citep{2003Quantum}
\begin{eqnarray}
\mathcal{L} & = & \int d^{3}\mathbf{x}\psi_{\alpha}^{\dagger}(t,\mathbf{x})\left(i\hbar\frac{\partial}{\partial t}+\frac{\hbar^{2}}{2m}\nabla^{2}\right)\psi_{\alpha}(t,\mathbf{x})\nonumber \\
 &  & -\frac{1}{2}\int d^{3}\mathbf{x}d^{3}\mathbf{x}^{\prime}\psi_{\alpha}^{\dagger}(t,\mathbf{x})\psi_{\beta}^{\dagger}(t,\mathbf{x}^{\prime})V_{\alpha\alpha^{\prime},\beta\beta^{\prime}}(\mathbf{x},\mathbf{x}^{\prime})\psi_{\beta^{\prime}}(t,\mathbf{x}^{\prime})\psi_{\alpha^{\prime}}(t,\mathbf{x}),\label{eq:lagrangian-1}
\end{eqnarray}
where $\alpha,\beta=\pm$ denote spin states, $V_{\alpha\alpha^{\prime},\beta\beta^{\prime}}(\mathbf{x},\mathbf{x}^{\prime})$
is the spin-dependent potential, and repeated indices imply a summation
if not explicitly stated. Let us consider the contact interaction
of the NJL type \citep{PhysRev.122.345,Nambu:1961fr}, 
\begin{equation}
\mathcal{L}_{\mathrm{int}}=-g_{0}\int d^{3}\mathbf{x}\left[\psi_{\alpha}^{\dagger}(t,\mathbf{x})\psi_{\alpha}(t,\mathbf{x})\right]^{2}-g_{\sigma}\sum_{i}\int d^{3}\mathbf{x}\left[\psi_{\alpha}^{\dagger}(t,\mathbf{x})\sigma_{\alpha\beta}^{i}\psi_{\beta}(t,\mathbf{x})\right]^{2},\label{eq:l-int-njl}
\end{equation}
which correpsonds to the potential in the form 
\begin{equation}
V_{\alpha\alpha^{\prime},\beta\beta^{\prime}}(\mathbf{x},\mathbf{x}^{\prime})=2\delta^{(3)}(\mathbf{x}-\mathbf{x}^{\prime})\left(g_{0}\delta_{\alpha\alpha^{\prime}}\delta_{\beta\beta^{\prime}}+g_{\sigma}\sum_{i}\sigma_{\alpha\alpha^{\prime}}^{i}\sigma_{\beta\beta^{\prime}}^{i}\right).\label{eq:njl-pot}
\end{equation}

The Lagrangian (\ref{eq:lagrangian-1}) with the interaction part
(\ref{eq:l-int-njl}) is invariant under the global SU(2) transformation
defined as 
\begin{eqnarray}
U(\theta) & = & \exp\left(-i\frac{1}{2}\boldsymbol{\theta}\cdot\boldsymbol{\sigma}\right),\nonumber \\
\psi_{\alpha}^{\prime} & = & U_{\alpha\beta}(\theta)\psi_{\beta},\nonumber \\
\psi_{\alpha}^{\prime\dagger} & = & \psi_{\beta}^{\dagger}U_{\beta\alpha}^{\dagger}(\theta),
\end{eqnarray}
which means the spin is conserved. The kinetic term of the Lagrangian
(\ref{eq:lagrangian-1}) and $g_{0}$ term of the interaction Lagrangian
(\ref{eq:l-int-njl}) are obviously invariant under the SU(2) transformation.
Let us look at the $g_{\sigma}$ term 
\begin{eqnarray}
\left[\psi_{\alpha}^{\prime\dagger}\sigma_{\alpha\beta}^{i}\psi_{\beta}^{\prime}\right]\left[\psi_{\alpha_{1}}^{\prime\dagger}\sigma_{\alpha_{1}\beta_{1}}^{i}\psi_{\beta_{1}}^{\prime}\right] & = & \left[\psi_{\gamma}^{\dagger}U_{\gamma\alpha}^{\dagger}\sigma_{\alpha\beta}^{i}U_{\beta\lambda}\psi_{\lambda}\right]\left[\psi_{\gamma_{1}}^{\dagger}U_{\gamma_{1}\alpha_{1}}^{\dagger}\sigma_{\alpha_{1}\beta_{1}}^{i}U_{\beta_{1}\lambda_{1}}\psi_{\lambda_{1}}\right]\nonumber \\
 & = & V_{ij}V_{ik}\left[\psi_{\gamma}^{\dagger}\sigma_{\gamma\lambda}^{j}\psi_{\lambda}\right]\left[\psi_{\gamma_{1}}^{\dagger}\sigma_{\gamma_{1}\lambda_{1}}^{k}\psi_{\lambda_{1}}\right]\nonumber \\
 & = & \left[\psi_{\gamma}^{\dagger}\sigma_{\gamma\lambda}^{j}\psi_{\lambda}\right]\left[\psi_{\gamma_{1}}^{\dagger}\sigma_{\gamma_{1}\lambda_{1}}^{j}\psi_{\lambda_{1}}\right],
\end{eqnarray}
where we have used $U^{\dagger}\sigma_{i}U=V_{ij}\sigma_{j}$ with
$V_{ij}$ denoting SO(3) matrices. Corresponding to the SU(2) invariance
of the Lagrangian, the Noether charge and current for spin are given
by 
\begin{eqnarray}
Q_{i}^{\mathrm{spin}} & = & \frac{\hbar}{2}\psi^{\dagger}\sigma_{i}\psi,\nonumber \\
J_{ij}^{\mathrm{spin}} & = & \frac{i\hbar^{2}}{4m}\left(\nabla_{j}\psi^{\dagger}\sigma_{i}\psi-\psi^{\dagger}\sigma_{i}\nabla_{j}\psi\right),
\end{eqnarray}
which satisfy the conservation equation 
\begin{equation}
\frac{\partial}{\partial t}Q_{i}^{\mathrm{spin}}+\nabla_{j}J_{ij}^{\mathrm{spin}}=0.
\end{equation}
Note that the Noether charge is a spin vector while the Noether current
is a tensor.

The fermion fields can be quantized as 
\begin{eqnarray}
\psi(x) & = & \int\frac{d^{3}\mathbf{p}}{(2\pi\hbar)^{3}}e^{-ip\cdot x/\hbar}\sum_{s=\pm}a(s,\mathbf{p})\chi(s),\nonumber \\
\psi^{\dagger}(x) & = & \int\frac{d^{3}\mathbf{p}}{(2\pi\hbar)^{3}}e^{ip\cdot x/\hbar}\sum_{s=\pm}a^{\dagger}(s,\mathbf{p})\chi^{\dagger}(s),\label{eq:quantum-field}
\end{eqnarray}
where $x\equiv(x_{0},\mathbf{x})\equiv(t,\mathbf{x})$, $p\equiv(p_{0},\mathbf{p})\equiv(\omega_{p},\mathbf{p})$
with $\omega_{p}=\mathbf{p}^{2}/(2m)$, $p\cdot x\equiv\omega_{p}t-\mathbf{p}\cdot\mathbf{x}$,
$a(s,\mathbf{p})$ and $a^{\dagger}(s,\mathbf{p})$ are annihilation
and creation operators associated with $\mathbf{p}$ and the spin
state $s$ respectively, and $\chi(s)$ is the spin state (Pauli spinor)
which satisfies $(\mathbf{n}\cdot\boldsymbol{\sigma})\chi(s)=s\chi(s)$
with $\mathbf{n}$ being the spin quantum direction $\mathbf{n}=(\sin\theta\cos\phi,\sin\theta\sin\phi,\cos\theta)$.
The anti-commutators of $a(s,\mathbf{p})$ and $a^{\dagger}(s,\mathbf{p})$
are given by 
\begin{eqnarray}
\left\{ a(s_{1},\mathbf{p}_{1}),a^{\dagger}(s_{2},\mathbf{p}_{2})\right\}  & = & (2\pi\hbar)^{3}\delta_{s_{1}s_{2}}\delta^{(3)}(\mathbf{p}_{1}-\mathbf{p}_{2}),\nonumber \\
\left\{ a(s_{1},\mathbf{p}_{1}),a(s_{2},\mathbf{p}_{2})\right\}  & = & \left\{ a^{\dagger}(s_{1},\mathbf{p}_{1}),a^{\dagger}(s_{2},\mathbf{p}_{2})\right\} =0,
\end{eqnarray}
which lead to equal-time anti-commutators for fermion fields 
\begin{eqnarray}
\left\{ \psi_{\alpha}(t,\mathbf{x}),\psi_{\beta}^{\dagger}(t,\mathbf{x}^{\prime})\right\}  & = & \delta_{\alpha\beta}\delta^{(3)}(\mathbf{x}-\mathbf{x}^{\prime}),\nonumber \\
\left\{ \psi_{\alpha}(t,\mathbf{x}),\psi_{\beta}(t,\mathbf{x}^{\prime})\right\}  & = & \left\{ \psi_{\alpha}^{\dagger}(t,\mathbf{x}),\psi_{\beta}^{\dagger}(t,\mathbf{x}^{\prime})\right\} =0.\label{eq:commutator}
\end{eqnarray}

Now we define the two-point Green function in the closed-time-path
(CTP) formalism \citep{Martin:1959jp,Keldysh:1964ud} (see, e.g.,
Refs. \citep{Chou:1984es,Blaizot:2001nr,Berges:2004yj,Crossley:2015evo}
for reviews) as 
\begin{equation}
G_{\alpha\beta}^{C}(x_{1},x_{2})=\left\langle T_{C}\left[\psi_{\alpha}(x_{1})\psi_{\beta}^{\dagger}(x_{2})\right]\right\rangle ,
\end{equation}
where $T_{C}$ denotes the time-ordered product on the CTP, and the
angular brackets denote averages weighted by the density operator
at the initial time $\rho(t_{0})$. Note that our definition for Green's
function is different from Ref. \citep{2003Quantum} without the additional
factor $i=\sqrt{-1}$. Depending on whether the two space-time points
are on the positive or negative time branch, there are four types
of two-point functions 
\begin{eqnarray}
G_{\alpha\beta}^{F}(x_{1},x_{2}) & = & G_{\alpha\beta}^{++}(x_{1},x_{2})=\left\langle T\psi_{\alpha}(x_{1})\psi_{\beta}^{\dagger}(x_{2})\right\rangle ,\nonumber \\
G_{\alpha\beta}^{\overline{F}}(x_{1},x_{2}) & = & G_{\alpha\beta}^{--}(x_{1},x_{2})=\left\langle T_{A}\psi_{\alpha}(x_{1})\psi_{\beta}^{\dagger}(x_{2})\right\rangle ,\nonumber \\
G_{\alpha\beta}^{<}(x_{1},x_{2}) & = & G_{\alpha\beta}^{+-}(x_{1},x_{2})=-\left\langle \psi_{\beta}^{\dagger}(x_{2})\psi_{\alpha}(x_{1})\right\rangle ,\nonumber \\
G_{\alpha\beta}^{>}(x_{1},x_{2}) & = & G_{\alpha\beta}^{-+}(x_{1},x_{2})=\left\langle \psi_{\alpha}(x_{1})\psi_{\beta}^{\dagger}(x_{2})\right\rangle ,\label{eq:def-green}
\end{eqnarray}
where $+/-$ stands for the positive/negative time branch respectively,
and $T$ and $T_{A}$ denote the time-ordered and reverse-time-ordered
product respectively. Note that only three out of four types of two-point
functions in (\ref{eq:def-green}) are independent due to the identity
\begin{equation}
G^{F}+G^{\overline{F}}=G^{<}+G^{>}.
\end{equation}
We can choose $(G^{F},G^{<},G^{>})$ as three independent two-point
functions. Equivalently we can also choose $(G^{R},G^{A},G^{S})$
as independent ones 
\begin{eqnarray}
G^{R} & = & G^{F}-G^{<}=\theta(t_{1}-t_{2})(G^{>}-G^{<}),\nonumber \\
G^{A} & = & G^{F}-G^{>}=\theta(t_{2}-t_{1})(G^{<}-G^{>}),\nonumber \\
G^{S} & = & G^{<}+G^{>},
\end{eqnarray}
where $G^{R}$ and $G^{A}$ are retarded and advanced two-point functions.

The Wigner function is the building-block of the quantum transport
theory since it is the quantum analogue of the particle distribution
in phase space. The Wigner function can be obtained by taking the
Fourier transformation with respect to the distance of two space-time
points in a two-point function. Then the Wigner functions for $G_{\alpha\beta}^{<}(x_{1},x_{2})$
and $G_{\alpha\beta}^{>}(x_{1},x_{2})$ are defined by 
\begin{eqnarray}
G_{\alpha\beta}^{<}(x,p) & = & \int d^{4}ye^{ip\cdot y/\hbar}G_{\alpha\beta}^{<}(x_{1},x_{2})\nonumber \\
 & = & -\int d^{4}ye^{ip\cdot y/\hbar}\left\langle \psi_{\beta}^{\dagger}\left(x-\frac{y}{2}\right)\psi_{\alpha}\left(x+\frac{y}{2}\right)\right\rangle ,\nonumber \\
G_{\alpha\beta}^{>}(x,p) & = & \int d^{4}ye^{ip\cdot y/\hbar}G_{\alpha\beta}^{>}(x_{1},x_{2})\nonumber \\
 & = & \int d^{4}ye^{ip\cdot y/\hbar}\left\langle \psi_{\alpha}\left(x+\frac{y}{2}\right)\psi_{\beta}^{\dagger}\left(x-\frac{y}{2}\right)\right\rangle ,\label{eq:def-wigner}
\end{eqnarray}
where $p\cdot y\equiv\omega_{p}y_{0}-\mathbf{p}\cdot\mathbf{y}$,
$x_{1}=x+y/2$ and $x_{2}=x-y/2$.


In non-equilibrium, we do not know $G_{\alpha\beta}^{<}(x,p)$ and
$G_{\alpha\beta}^{>}(x,p)$ exactly due to unknown ensemble averages.
However, using (\ref{eq:quantum-field}), we can make an ansatz for
their forms in a power expansion of $\hbar$. Inserting (\ref{eq:quantum-field})
to (\ref{eq:def-wigner}), the leading order contributions to the
Wigner functions have the form 
\begin{eqnarray}
G_{\alpha\beta}^{<(0)}(x,p) & = & -(2\pi\hbar)\delta\left(p_{0}-\frac{\mathbf{p}^{2}}{2m}\right)\sum_{s_{1},s_{2}=\pm}\chi_{\alpha}(s_{1})\chi_{\beta}^{\dagger}(s_{2})f_{s_{1}s_{2}}^{(0)}(x,\mathbf{p}),\nonumber \\
G_{\alpha\beta}^{>(0)}(x,p) & = & (2\pi\hbar)\delta\left(p_{0}-\frac{\mathbf{p}^{2}}{2m}\right)\sum_{s_{1},s_{2}=\pm}\chi_{\alpha}(s_{1})\chi_{\beta}^{\dagger}(s_{2})\left[\delta_{s_{1}s_{2}}-f_{s_{1}s_{2}}^{(0)}(x,\mathbf{p})\right],\label{eq:g0}
\end{eqnarray}
where $p\equiv(p_{0},\mathbf{p})\equiv(\omega,\mathbf{p})$ with $p_{0}$
or $\omega$ being an independent variable, and the matrix valued
spin-dependent distribution (MVSD) at $\mathcal{O}(\hbar^{0})$ is
defined as \citep{Sheng:2021kfc}
\begin{eqnarray}
f_{s_{1}s_{2}}^{(0)}(x,\mathbf{p}) & = & \int\frac{d^{3}\mathbf{q}}{(2\pi\hbar)^{3}}\exp\left[i\frac{1}{\hbar}\left(-\frac{\mathbf{p}\cdot\mathbf{q}}{m}t+\mathbf{q}\cdot\mathbf{x}\right)\right]\nonumber \\
 &  & \times\left\langle a^{\dagger}\left(s_{2},\mathbf{p}-\frac{\mathbf{q}}{2}\right)a\left(s_{1},\mathbf{p}+\frac{\mathbf{q}}{2}\right)\right\rangle .
\end{eqnarray}
One can check that $G^{\lessgtr(0)}(x,p)$ are Hermitian matrices
because $f_{s_{1}s_{2}}^{(0)*}=f_{s_{2}s_{1}}^{(0)}$, i.e., $f^{(0)}$
is a Hermitian matrix in spin space. The first order contributions
at $\mathcal{O}(\hbar^{1})$ are assumed to have the form
\begin{eqnarray}
\hbar G_{\alpha\beta}^{<(1)}(x,p) & = & \hbar G_{\alpha\beta}^{>(1)}(x,p)\nonumber \\
 & = & -\hbar(2\pi\hbar)\delta\left(p_{0}-\frac{\mathbf{p}^{2}}{2m}\right)\sum_{s_{1},s_{2}=\pm}\chi_{\alpha}(s_{1})\chi_{\beta}^{\dagger}(s_{2})f_{s_{1}s_{2}}^{(1)}(x,\mathbf{p}),\label{eq:g1}
\end{eqnarray}
where $f_{s_{1}s_{2}}^{(1)}(x,\mathbf{p})$ is unknown and can be
determined by solving the evolution equations. We assume $f_{s_{1}s_{2}}^{(1)*}=f_{s_{2}s_{1}}^{(1)}$,
which means $f^{(1)}$ and then $G^{\lessgtr(1)}(x,p)$ are Hermitian
matrices in spin space. Note that in Eq. (\ref{eq:g1}) we only include
on-shell contributions to $G^{\lessgtr(1)}(x,p)$ which are proportional
to $\delta\left[p_{0}-\mathbf{p}^{2}/(2m)\right]$. In principle there
are off-shell contributions which are proportional to $\delta^{\prime}\left[p_{0}-\mathbf{p}^{2}/(2m)\right]$.
We will address off-shell contributions in Section \ref{sec:off-shell-part}
separately.

Combining (\ref{eq:g0}) and (\ref{eq:g1}) we obtain $G^{\lessgtr}(x,p)$
up to $\mathcal{O}(\hbar)$ 
\begin{eqnarray}
G^{<}(x,p) & = & G^{<(0)}(x,p)+\hbar G^{<(1)}(x,p)+\mathcal{O}(\hbar^{2})\nonumber \\
 & = & -(2\pi\hbar)\delta\left(p_{0}-\frac{\mathbf{p}^{2}}{2m}\right)\sum_{s_{1},s_{2}=\pm}\chi_{\alpha}(s_{1})\chi_{\beta}^{\dagger}(s_{2})f_{s_{1}s_{2}}(x,\mathbf{p}),\nonumber \\
G^{>}(x,p) & = & G^{>(0)}(x,p)+\hbar G^{>(1)}(x,p)+\mathcal{O}(\hbar^{2})\nonumber \\
 & = & (2\pi\hbar)\delta\left(p_{0}-\frac{\mathbf{p}^{2}}{2m}\right)\sum_{s_{1},s_{2}=\pm}\chi_{\alpha}(s_{1})\chi_{\beta}^{\dagger}(s_{2})\left[\delta_{s_{1}s_{2}}-f_{s_{1}s_{2}}(x,\mathbf{p})\right].\label{eq:on-shell-g}
\end{eqnarray}
Here the MVSD $f_{s_{1}s_{2}}(x,\mathbf{p})$ is given by 
\begin{eqnarray}
f_{s_{1}s_{2}}(x,\mathbf{p}) & = & f_{s_{1}s_{2}}^{(0)}(x,\mathbf{p})+\hbar f_{s_{1}s_{2}}^{(1)}(x,\mathbf{p})+\mathcal{O}(\hbar^{2})\nonumber \\
 & = & \delta_{s_{1}s_{2}}\overline{f}(x,\mathbf{p})+\boldsymbol{\tau}_{s_{1}s_{2}}\cdot\mathbf{t}(x,\mathbf{p})+\mathcal{O}(\hbar^{2}),\label{eq:mvsd-f-s}
\end{eqnarray}
where $\boldsymbol{\tau}=(\tau_{1},\tau_{2},\tau_{3})$ and $\mathbf{t}(x,\mathbf{p})=[t_{1}(x,\mathbf{p}),t_{2}(x,\mathbf{p}),t_{3}(x,\mathbf{p})]$
denote Pauli matrices and the polarization vector in $(s_{1},s_{2})$
space respectively. Note that $G^{\lessgtr}(x,p)$ are Hermitian.
For convenience we can write $G^{\lessgtr}(x,p)$ as 
\begin{equation}
G^{\lessgtr}(x,p)=-(2\pi\hbar)\delta\left(p_{0}-\frac{\mathbf{p}^{2}}{2m}\right)g^{\lessgtr}(x,p),
\end{equation}
where $g^{\lessgtr}(x,p)$ denotes the part of $G^{\lessgtr}(x,p)$
without the delta function and it can be decomposed in terms of 1
and Pauli matrices $\boldsymbol{\sigma}=(\sigma_{1},\sigma_{2},\sigma_{3})$
in spinor's space, 
\begin{eqnarray}
g^{<}(x,p) & \equiv & \sum_{s_{1},s_{2}=\pm}\chi(s_{1})\chi^{\dagger}(s_{2})f_{s_{1}s_{2}}(x,\mathbf{p})\nonumber \\
 & = & \overline{f}(x,\mathbf{p})+\boldsymbol{\sigma}\cdot\mathbf{S}(x,\mathbf{p}),\nonumber \\
g^{>}(x,p) & \equiv & -\sum_{s_{1},s_{2}=\pm}\chi(s_{1})\chi^{\dagger}(s_{2})\left[\delta_{s_{1}s_{2}}-f_{s_{1}s_{2}}(x,\mathbf{p})\right]\nonumber \\
 & = & \overline{f}(x,\mathbf{p})+\boldsymbol{\sigma}\cdot\mathbf{S}(x,\mathbf{p})-1.\label{eq:def-f-s}
\end{eqnarray}
These components can be extracted by taking traces 
\begin{eqnarray}
\overline{f}(x,\mathbf{p}) & = & \frac{1}{2}\mathrm{Tr}\left[g^{<}(x,p)\right],\nonumber \\
\mathbf{S}(x,\mathbf{p}) & = & \frac{1}{2}\mathrm{Tr}\left[\boldsymbol{\sigma}g^{<}(x,p)\right].
\end{eqnarray}
So $\mathbf{S}(x,\mathbf{p})$ is given by 
\begin{eqnarray}
\mathbf{S}(x,\mathbf{p}) & = & \frac{1}{2}\sum_{s_{1},s_{2}=\pm}\chi^{\dagger}(s_{2})\boldsymbol{\sigma}\chi(s_{1})f_{s_{1}s_{2}}(x,\mathbf{p})\nonumber \\
 & = & \frac{1}{2}\sum_{s_{1},s_{2}=\pm}(\mathbf{n}_{1}\tau_{1}+\mathbf{n}_{2}\tau_{2}+\mathbf{n}_{3}\tau_{3})_{s_{2}s_{1}}f_{s_{1}s_{2}}(x,\mathbf{p})\nonumber \\
 & = & \mathbf{n}_{i}t_{i}(x,\mathbf{p}),
\end{eqnarray}
where we have used 
\begin{eqnarray}
\chi^{\dagger}(s_{2})\boldsymbol{\sigma}\chi(s_{1}) & = & (\mathbf{n}_{1}\tau_{1}+\mathbf{n}_{2}\tau_{2}+\mathbf{n}_{3}\tau_{3})_{s_{2}s_{1}}\nonumber \\
 & = & \left(\begin{array}{cc}
\mathbf{n}_{3} & \mathbf{n}_{1}-i\mathbf{n}_{2}\\
\mathbf{n}_{1}+i\mathbf{n}_{2} & -\mathbf{n}_{3}
\end{array}\right)_{s_{2}s_{1}}.
\end{eqnarray}
Here $(\mathbf{n}_{1},\mathbf{n}_{2},\mathbf{n}_{3})$ are three basis
vectors given by 
\begin{eqnarray}
\mathbf{n}_{1} & = & \left(\cos\phi\cos\theta,\sin\phi\cos\theta,-\sin\theta\right),\nonumber \\
\mathbf{n}_{2} & = & \left(-\sin\phi,\cos\phi,0\right),\nonumber \\
\mathbf{n}_{3} & = & \left(\sin\theta\cos\phi,\sin\theta\sin\phi,\cos\theta\right).\label{eq:n1n2n3}
\end{eqnarray}
Note that $\mathbf{n}_{3}=\mathbf{n}$ is the direction of the spin
quantization.

\section{Kadanoff-Baym equation}

\label{sec:Kadanoff-Baym-equation}The time evolution of a many-body
quantum system in non-equilibrium is described by the Kadanoff-Baym
(KB) equation \citep{Kadanoff2018QuantumSM} {[}see, e.g., Ref. \citep{Cassing:2008nn}
for a review{]}. The KB equations can be derived from the Dyson-Schwinger
equations for two-point Green functions on the CTP 
\begin{eqnarray}
-i\left(i\hbar\frac{\partial}{\partial t_{1}}+\frac{\hbar^{2}}{2m}\nabla_{x1}^{2}\right)G_{C}(x_{1},x_{2}) & = & \hbar\delta_{C}^{(4)}(x_{1}-x_{2})+\hbar\int_{C}d^{4}x^{\prime}\Sigma_{C}(x_{1},x^{\prime})G_{C}(x^{\prime},x_{2}),\nonumber \\
-iG_{C}(x_{1},x_{2})\left(-i\hbar\frac{\overleftarrow{\partial}}{\partial t_{2}}+\frac{\hbar^{2}}{2m}\overleftarrow{\nabla}_{x2}^{2}\right) & = & \hbar\delta_{C}^{(4)}(x_{1}-x_{2})+\hbar\int_{C}d^{4}x^{\prime}G_{C}(x_{1},x^{\prime})\Sigma_{C}(x^{\prime},x_{2}),\label{eq:dyson-equation}
\end{eqnarray}
where the index $C$ stands for the CTP, $\delta_{C}^{(4)}(x_{1}-x_{2})$
and $\Sigma_{C}(x_{1},x^{\prime})$ are the delta-function and self-energy
on the CTP respectively. Note that $G_{C}$ and $\Sigma_{C}$ are
$2\times2$ matrices in spinor space. In the case that $(t_{1},t_{2})$
are on ($+,-$) time branch, Eq. (\ref{eq:dyson-equation}) can be
put into the conventional coordinate form 
\begin{eqnarray}
-i\left(i\hbar\frac{\partial}{\partial t_{1}}+\frac{\hbar^{2}}{2m}\nabla_{x1}^{2}\right)G^{<}(x_{1},x_{2}) & = & \hbar\int d^{4}x^{\prime}\left[\Sigma^{R}(x_{1},x^{\prime})G^{<}(x^{\prime},x_{2})\right.\nonumber \\
 &  & \left.+\Sigma^{<}(x_{1},x^{\prime})G^{A}(x^{\prime},x_{2})\right],\label{eq:kb-g-left}\\
-i\left(-i\hbar\frac{\partial}{\partial t_{2}}+\frac{\hbar^{2}}{2m}\nabla_{x2}^{2}\right)G^{<}(x_{1},x_{2}) & = & \hbar\int d^{4}x^{\prime}\left[G^{R}(x_{1},x^{\prime})\Sigma^{<}(x^{\prime},x_{2})\right.\nonumber \\
 &  & \left.+G^{<}(x_{1},x^{\prime})\Sigma^{A}(x^{\prime},x_{2})\right],\label{eq:kb-g-less-g-larger}
\end{eqnarray}
where $\Sigma^{R}$ and $\Sigma^{A}$ are the retarded and advanced
self-energy respectively. Performing the Wigner transform for above
equations, we obtain
\begin{equation}
\left(i\hbar\frac{1}{2}\partial_{t}+i\frac{\hbar}{2m}\mathbf{p}\cdot\nabla_{x}+p_{0}-\frac{\mathbf{p}^{2}}{2m}+\frac{\hbar^{2}}{8m}\nabla_{x}^{2}\right)G^{<}(x,p)=I_{\mathrm{coll}},\label{eq:left-kb}
\end{equation}
 and 
\begin{equation}
\left(-i\hbar\frac{1}{2}\partial_{t}-i\frac{\hbar}{2m}\mathbf{p}\cdot\nabla_{x}+p_{0}-\frac{\mathbf{p}^{2}}{2m}+\frac{\hbar^{2}}{8m}\nabla_{x}^{2}\right)G^{<}(x,p)=I_{\mathrm{coll}}^{\dagger},\label{eq:right-kb}
\end{equation}
where $I_{\mathrm{coll}}$ and $I_{\mathrm{coll}}^{\dagger}$ are
given by 
\begin{eqnarray}
I_{\mathrm{coll}} & = & i\hbar\left[\Sigma^{R}(x,p)G^{<}(x,p)+\Sigma^{<}(x,p)G^{A}(x,p)\right]\nonumber \\
 &  & +\frac{1}{2}\hbar^{2}\left[\left\{ \Sigma^{R}(x,p),G^{<}(x,p)\right\} _{P.B.}+\left\{ \Sigma^{<}(x,p),G^{A}(x,p)\right\} _{P.B.}\right],\label{eq:i-coll}\\
I_{\mathrm{coll}}^{\dagger} & = & i\hbar\left[G^{R}(x,p)\Sigma^{<}(x,p)+G^{<}(x,p)\Sigma^{A}(x,p)\right]\nonumber \\
 &  & +\frac{1}{2}\hbar^{2}\left[\left\{ G^{R}(x,p),\Sigma^{<}(x,p)\right\} _{P.B.}+\left\{ G^{<}(x,p),\Sigma^{A}(x,p)\right\} _{P.B.}\right].\label{eq:i-coll-hermite}
\end{eqnarray}
Here the Poisson bracket of two matrices is defined as 
\begin{equation}
\left\{ A,B\right\} _{\mathrm{PB}}\equiv\partial_{t}A\partial_{p_{0}}B-\partial_{p_{0}}A\partial_{t}B-\left(\nabla_{x}A\cdot\nabla_{p}B-\nabla_{p}A\cdot\nabla_{x}B\right).\label{eq:poisson-bra}
\end{equation}
We see that $I_{\mathrm{coll}}^{\dagger}$ can be obtained by interchange
of $\Sigma$ and $G$ from $I_{\mathrm{coll}}$, and vice versa. With
the relations for $O=G,\Sigma$, 
\begin{equation}
[O^{R}]^{\dagger}=-O^{A},\;\;\left[O^{\lessgtr}\right]^{\dagger}=\left[O^{\lessgtr}\right],
\end{equation}
one can check that $I_{\mathrm{coll}}^{\dagger}$ is really the Hermitian
conjugate of $I_{\mathrm{coll}}$. Taking the sum and difference of
Eq. (\ref{eq:i-coll}) and (\ref{eq:i-coll-hermite}), we obtain an
equation for the dispersion relation or the on-shell equation
\begin{equation}
\left(p_{0}-\frac{\mathbf{p}^{2}}{2m}+\frac{\hbar^{2}}{8m}\nabla_{x}^{2}\right)G^{<}(x,p)=\frac{1}{2}\left(I_{\mathrm{coll}}+I_{\mathrm{coll}}^{\dagger}\right),\label{eq:on-shell}
\end{equation}
and the evolution equation 
\begin{equation}
\hbar\left(\partial_{t}+\frac{\mathbf{p}}{m}\cdot\nabla_{x}\right)G^{<}(x,p)=-i\left(I_{\mathrm{coll}}-I_{\mathrm{coll}}^{\dagger}\right).\label{eq:kb-evo}
\end{equation}
Equations (\ref{eq:on-shell},\ref{eq:kb-evo}) are one of the main
results in this paper and the starting point for the derivation of
Boltzmann equations.


In the quasi-particle picture, the retarded and advanced self-energies
and two-point functions can be approximated as 
\begin{eqnarray}
O^{R/A}(x,p) & = & \frac{1}{2\pi i}\int dk_{0}\frac{1}{k_{0}-p_{0}\mp i\epsilon}\left[O^{>}(x,k_{0},\mathbf{p})-O^{<}(x,k_{0},\mathbf{p})\right]\nonumber \\
 & = & \pm\frac{1}{2}\left[O^{>}(x,p)-O^{<}(x,p)\right]+O_{\mathrm{pr}}^{R/A},\label{eq:ret-adv-approx}
\end{eqnarray}
where $O\equiv G,\Sigma$, and the principal part $O_{\mathrm{pr}}^{R/A}$
is related to off-shell effects that modify the dispersion relation
of the quasi-particle which we will not consider in this paper. Using
(\ref{eq:ret-adv-approx}), the collision terms (\ref{eq:i-coll})
and (\ref{eq:i-coll-hermite}) can be simplified as
\begin{eqnarray}
I_{\mathrm{coll}} & = & \frac{1}{2}i\hbar\left[\Sigma^{>}(x,p)G^{<}(x,p)-\Sigma^{<}(x,p)G^{>}(x,p)\right]\nonumber \\
 &  & +\frac{1}{4}\hbar^{2}\left[\left\{ \Sigma^{>}(x,p),G^{<}(x,p)\right\} _{P.B.}-\left\{ \Sigma^{<}(x,p),G^{>}(x,p)\right\} _{P.B.}\right],\label{eq:coll-on-shell}
\end{eqnarray}
and 
\begin{eqnarray}
I_{\mathrm{coll}}^{\dagger} & = & \frac{1}{2}i\hbar\left[G^{>}(x,p)\Sigma^{<}(x,p)-G^{<}(x,p)\Sigma^{>}(x,p)\right]\nonumber \\
 &  & +\frac{1}{4}\hbar^{2}\left[\left\{ G^{>}(x,p),\Sigma^{<}(x,p)\right\} _{P.B.}-\left\{ G^{<}(x,p),\Sigma^{>}(x,p)\right\} _{P.B.}\right].\label{eq:coll-on-shell-conj}
\end{eqnarray}
With collision terms (\ref{eq:coll-on-shell}) and (\ref{eq:coll-on-shell-conj}),
Eq. (\ref{eq:kb-evo}) is our starting point for the derivation of
Boltzmann equations.

The Boltzmann equations for $\overline{f}(x,\mathbf{p})$ and $\mathbf{S}(x,p)$
can be obtained by taking a trace of Eq. (\ref{eq:kb-evo}) and a
trace of Eq. (\ref{eq:kb-evo}) multiplied by $\boldsymbol{\sigma}$
as 
\begin{eqnarray}
\hbar\left(\partial_{t}+\frac{1}{m}\mathbf{p}\cdot\nabla_{x}\right)\mathrm{Tr}\left(G^{<}\right) & = & 2\mathrm{Im}\mathrm{Tr}\left(I_{\mathrm{coll}}\right),\label{eq:boltzman-f}\\
\hbar\left(\partial_{t}+\frac{1}{m}\mathbf{p}\cdot\nabla_{x}\right)\mathrm{Tr}\left(\boldsymbol{\sigma}G^{<}\right) & = & 2\mathrm{Im}\mathrm{Tr}\left(\boldsymbol{\sigma}I_{\mathrm{coll}}\right).\label{eq:boltzman-s}
\end{eqnarray}
From (\ref{eq:on-shell-g}) we know that there is an on-shell delta-function
$\delta\left[p_{0}-\mathbf{p}^{2}/(2m)\right]$ in both sides of above
equations, therefore one can drop these delta-functions and derive
equations for $\overline{f}(x,\mathbf{p})$ and $\mathbf{S}(x,p)$
without delta-functions.


\section{Spin boltzmann equations: on-shell parts}

\label{sec:NJL-on-shell}
\begin{figure}
\caption{\label{fig:feynman}Feynman diagrams for $\Sigma^{>}(x,p)$ {[}figure
(a) and (c){]} and $\Sigma^{<}(x,p)$ {[}figure (b) and (d){]}. In
the NJL model wavy lines attached with two solid circles represent
vertices $2g_{c}\Gamma^{(c)}$.  }

\includegraphics[scale=0.4]{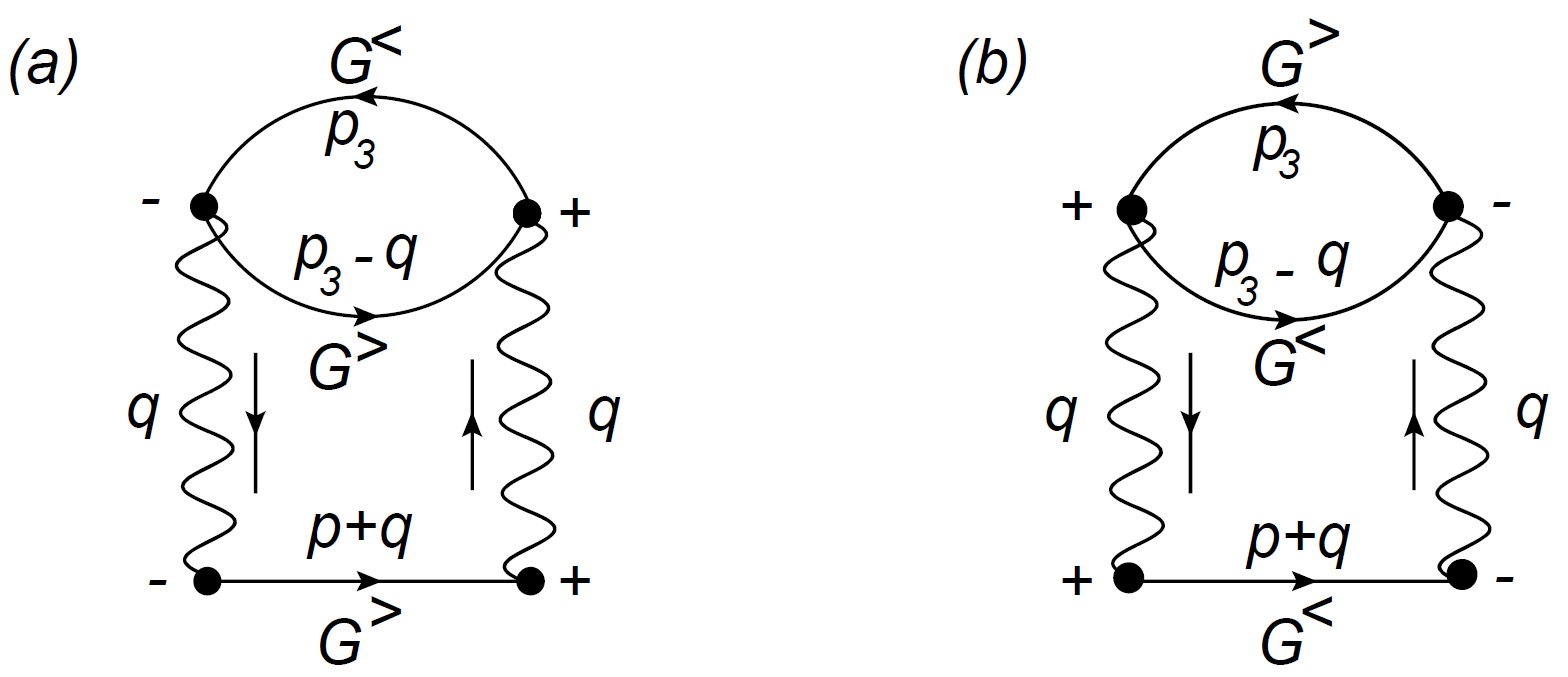}

\includegraphics[scale=0.4]{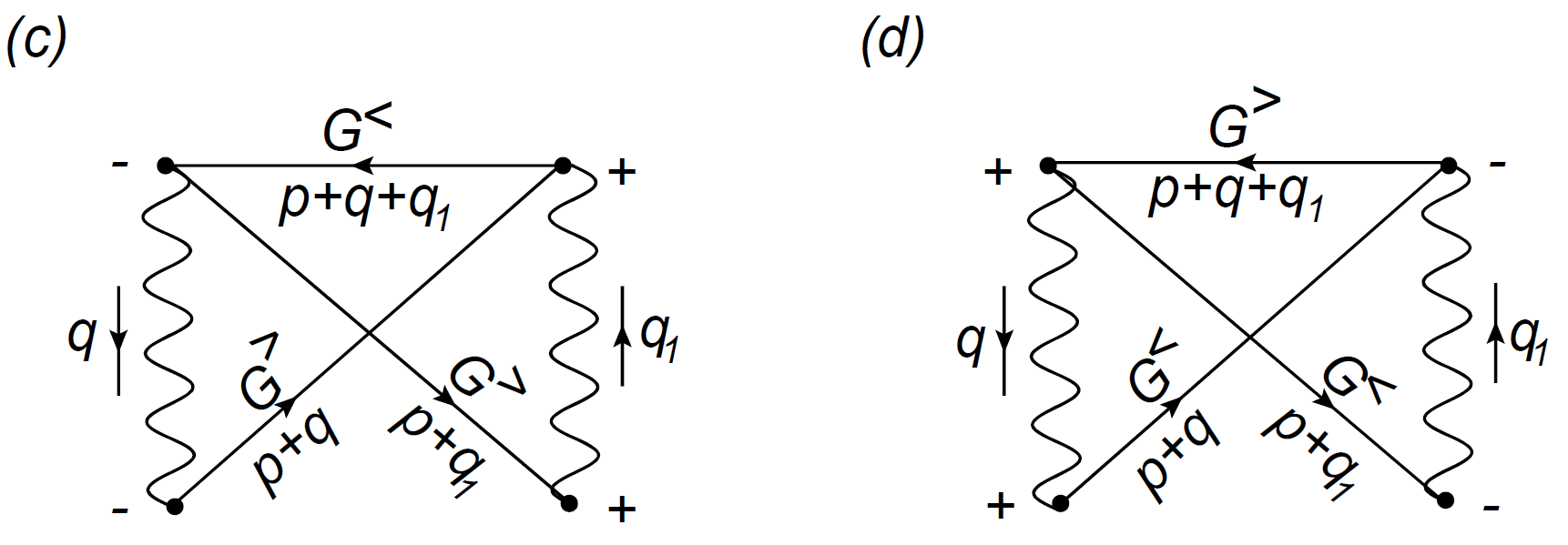}
\end{figure}

In this section we consider the contact interaction
of the NJL type, one of the simplest cases for interaction. The interaction
Lagrangian is given in (\ref{eq:l-int-njl}). The collision term $I_{\mathrm{coll}}$
in (\ref{eq:coll-on-shell}) and $I_{\mathrm{coll}}^{\dagger}$ in
(\ref{eq:coll-on-shell-conj}) depend on the self-energy $\Sigma^{>}$
and $\Sigma^{<}$ whose Feynman diagrams are shown in Fig. \ref{fig:feynman}.
The self-energies can be written as 
\begin{eqnarray}
\Sigma^{>}(x,p) & = & 4\sum_{c1,c2}g_{c1}g_{c2}\int\frac{d^{4}p_{1}}{(2\pi\hbar)^{4}}\frac{d^{4}p_{2}}{(2\pi\hbar)^{4}}\frac{d^{4}p_{3}}{(2\pi\hbar)^{4}}(2\pi\hbar)^{4}\delta^{(4)}(p+p_{3}-p_{1}-p_{2})\nonumber \\
 &  & \times\mathrm{Tr}\left[\Gamma^{(c2)}G^{>}(p_{1})\Gamma^{(c1)}G^{<}(p_{3})\right]\Gamma^{(c2)}G^{>}(p_{2})\Gamma^{(c1)}\nonumber \\
 &  & -4\sum_{c1,c2}g_{c1}g_{c2}\int\frac{d^{4}p_{1}}{(2\pi\hbar)^{4}}\frac{d^{4}p_{2}}{(2\pi\hbar)^{4}}\frac{d^{4}p_{3}}{(2\pi\hbar)^{4}}(2\pi\hbar)^{4}\delta^{(4)}(p+p_{3}-p_{1}-p_{2})\nonumber \\
 &  & \times\Gamma^{(c2)}G^{>}(p_{1})\Gamma^{(c1)}G^{<}(p_{3})\Gamma^{(c2)}G^{>}(p_{2})\Gamma^{(c1)},\nonumber \\
\Sigma^{<}(x,p) & = & \Sigma^{>}[G^{>}\leftrightarrow G^{<}],\label{eq:self-en-1}
\end{eqnarray}
where $g_{c1}$ and $g_{c2}$ can be $g_{0}$ or $g_{\sigma}$, and
we have suppressed the $x$ dependence of $G^{<}$ and $G^{>}$. If
$g_{c}$ is $g_{\sigma}$, a summation over $i$ for $\Gamma^{(\sigma)}=\sigma_{i}$
is implied.

\subsection{Leading order}

Using Eqs. (\ref{eq:on-shell-g}), (\ref{eq:self-en-1}) and (\ref{eq:coll-on-shell})
in Eqs.(\ref{eq:boltzman-f}) and (\ref{eq:boltzman-s}) and performing
an integration of $p_{0}$ over 0 to $+\infty$, we obtain the Boltzmann
equations for the scalar and polarization part of the distribution
at the leading order 
\begin{eqnarray}
\hbar\left(\partial_{t}+\frac{1}{m}\mathbf{p}\cdot\nabla_{x}\right)\overline{f}_{p}^{(0)} & = & 4\hbar(g_{0}-3g_{\sigma})^{2}\int\frac{d^{3}\mathbf{p}_{1}}{(2\pi\hbar)^{3}}\frac{d^{3}\mathbf{p}_{2}}{(2\pi\hbar)^{3}}\frac{d^{3}\mathbf{p}_{3}}{(2\pi\hbar)^{3}}(2\pi\hbar)^{4}\delta^{(4)}(p+p_{3}-p_{2}-p_{1})\nonumber \\
 &  & \times\left[\overline{f}_{1}^{(0)}\overline{f}_{2}^{(0)}(1-\overline{f}_{3}^{(0)})(1-\overline{f}_{p}^{(0)})-(1-\overline{f}_{1}^{(0)})(1-\overline{f}_{2}^{(0)})\overline{f}_{3}^{(0)}\overline{f}_{p}^{(0)}\right.\nonumber \\
 &  & \left.-(1-\overline{f}_{3}^{(0)}-\overline{f}_{p}^{(0)})\mathbf{S}_{1}^{(0)}\cdot\mathbf{S}_{2}^{(0)}+(1-\overline{f}_{1}^{(0)}-\overline{f}_{2}^{(0)})\mathbf{S}_{3}^{(0)}\cdot\mathbf{S}_{p}^{(0)}\right],\label{eq:leading-NJL-boltz-f}\\
\hbar\left(\partial_{t}+\frac{1}{m}\mathbf{p}\cdot\nabla_{x}\right)\mathbf{S}_{p}^{a,(0)} & = & 4\hbar(g_{0}-3g_{\sigma})^{2}\int\frac{d^{3}\mathbf{p}_{1}}{(2\pi\hbar)^{3}}\frac{d^{3}\mathbf{p}_{2}}{(2\pi\hbar)^{3}}\frac{d^{3}\mathbf{p}_{3}}{(2\pi\hbar)^{3}}(2\pi\hbar)^{4}\delta^{(4)}(p+p_{3}-p_{2}-p_{1})\nonumber \\
 &  & \times\left\{ \left[(1-\overline{f}_{p}^{(0)})\overline{f}_{1}^{(0)}\overline{f}_{2}^{(0)}+\overline{f}_{p}^{(0)}(1-\overline{f}_{1}^{(0)})(1-\overline{f}_{2}^{(0)})\right]\mathbf{S}_{3}^{a,(0)}\right.\nonumber \\
 &  & -\left[(1-\overline{f}_{3}^{(0)})\overline{f}_{1}^{(0)}\overline{f}_{2}^{(0)}+\overline{f}_{3}^{(0)}(1-\overline{f}_{1}^{(0)})(1-\overline{f}_{2}^{(0)})\right]\mathbf{S}_{p}^{a,(0)}\nonumber \\
 &  & \left.+\mathbf{S}_{1}^{(0)}\cdot\mathbf{S}_{2}^{(0)}(\mathbf{S}_{p}^{a,(0)}-\mathbf{S}_{3}^{a,(0)})\right\} ,\label{eq:leading-NJL-pol-s}
\end{eqnarray}
where $\overline{f}_{i}\equiv\overline{f}(x,\mathbf{p}_{i})$, $\overline{f}_{p}\equiv\overline{f}(x,\mathbf{p})$,
$\mathbf{S}_{i}\equiv\mathbf{S}(x,\mathbf{p}_{i})$, $\mathbf{S}_{p}\equiv\mathbf{S}(x,\mathbf{p})$,
and the index '(0)' denotes the leading order. If the system has no
polarization at the leading order, i.e. $\mathbf{S}_{i}^{(0)}=0$
for $i=1,2,3,p$, Eq. (\ref{eq:leading-NJL-boltz-f}) is reduced to
the conventional Boltzmann equation for $\overline{f}$, with Eq.
(\ref{eq:leading-NJL-pol-s}) for $\mathbf{S}^{a}$ being trivially
satisfied (both sides are vanishing). If we assume polarized distributions
are much smaller in magnitude than unpolarized ones, i.e. $\left|\mathbf{S}_{i}^{(0)}\right|\ll\overline{f}_{j}$
for $i,j=1,2,3,p$, then we can neglect quadratic terms of polarized
distributions in Eq. (\ref{eq:leading-NJL-boltz-f}) relative to terms
with only unpolarized distributions and neglect cubic terms of polarized
distributions in Eq. (\ref{eq:leading-NJL-pol-s}) relative to linear
terms. In this case the vanishing of the collision term in Eq. (\ref{eq:leading-NJL-boltz-f})
gives the equilibrium condition for unpolarized distributions 
\begin{equation}
\overline{f}_{1}^{(0)}\overline{f}_{2}^{(0)}(1-\overline{f}_{3}^{(0)})(1-\overline{f}_{p}^{(0)})=(1-\overline{f}_{1}^{(0)})(1-\overline{f}_{2}^{(0)})\overline{f}_{3}^{(0)}\overline{f}_{p}^{(0)}.\label{eq:cond-f}
\end{equation}
Similarly we can also obtain the equilibrium condition for unpolarized
distributions from the vanishing of the collision term in Eq. (\ref{eq:leading-NJL-pol-s})
\begin{align}
 & \left[\left(1-\overline{f}_{p}^{(0)}\right)\overline{f}_{1}^{(0)}\overline{f}_{2}^{(0)}+\overline{f}_{p}^{(0)}\left(1-\overline{f}_{1}^{(0)}\right)\left(1-\overline{f}_{2}^{(0)}\right)\right]\mathbf{S}_{3}^{a,(0)}\nonumber \\
 & =\left[\left(1-\overline{f}_{3}^{(0)}\right)\overline{f}_{1}^{(0)}\overline{f}_{2}^{(0)}+\overline{f}_{3}^{(0)}\left(1-\overline{f}_{1}^{(0)}\right)\left(1-\overline{f}_{2}^{(0)}\right)\right]\mathbf{S}_{p}^{a,(0)},
\end{align}
which leads to 
\begin{equation}
\frac{\mathbf{S}_{3}^{(0)}}{\left(1-\overline{f}_{3}^{(0)}\right)\overline{f}_{3}^{(0)}}=\frac{\mathbf{S}_{p}^{(0)}}{\left(1-\overline{f}_{p}^{(0)}\right)\overline{f}_{p}^{(0)}},\label{eq:eq-cond-s}
\end{equation}
using the equilibrium condition (\ref{eq:cond-f}). The equilibrium
condition (\ref{eq:cond-f}) implies that $\overline{f}_{p}^{(0)}$
follows the Fermi-Dirac distribution
\begin{equation}
\overline{f}_{p}^{(0)}=\frac{1}{\exp\left[\beta(\omega_{p}-\mu)\right]+1}.
\end{equation}
If we assume the ratio in Eq. (\ref{eq:eq-cond-s}) is a constant
vector $\mathbf{c}$ which is related to the spin potential $\boldsymbol{\mu}_{\mathrm{spin}}$,
then at the leading order the MVSD in Eq. (\ref{eq:mvsd-f-s}) has
the form 
\begin{equation}
f_{s_{1}s_{2}}^{(0)}(x,\mathbf{p})=\delta_{s_{1}s_{2}}\overline{f}_{p}^{(0)}+\overline{f}_{p}^{(0)}\left(1-\overline{f}_{p}^{(0)}\right)\boldsymbol{\tau}_{s_{1}s_{2}}\cdot\boldsymbol{\mu}_{\mathrm{spin}}+\mathcal{O}(\hbar),
\end{equation}
where the components of $\boldsymbol{\mu}_{\mathrm{spin}}$ are $\mu_{\mathrm{spin}}^{i}=\mathbf{n}_{i}\cdot\mathbf{c}$
($i=1,2,3$) with three directions $(\mathbf{n}_{1},\mathbf{n}_{2},\mathbf{n}_{3})$
being given by Eq. (\ref{eq:n1n2n3}). Since we have assumed $\left|\mathbf{S}_{i}^{(0)}\right|\ll\overline{f}_{j}$,
i.e. $|\boldsymbol{\mu}_{\mathrm{spin}}|\ll1$, $f_{s_{1}s_{2}}^{(0)}$
can be put into an approximated matrix form 
\begin{equation}
f^{(0)}(x,\mathbf{p})\approx\frac{1}{\exp\left[\beta(\omega_{p}-\mu)-\boldsymbol{\tau}\cdot\boldsymbol{\mu}_{\mathrm{spin}}\right]+1}.\label{eq:equil-spin}
\end{equation}
Here $\boldsymbol{\tau}=(\tau_{1},\tau_{2},\tau_{3})$ are Pauli matrices
in spin space. The MVSD in (\ref{eq:equil-spin}) is the equilibrium
distribution for fermions with spin degrees of freedom.

\subsection{Next-to-leading order}

At next to leading order, the Boltzmann equation for the unpolarized
distribution reads 
\begin{equation}
\hbar(\partial_{t}+\frac{1}{m}\mathbf{p}\cdot\nabla_{x})\overline{f}_{p}^{(1)}=I_{\mathrm{qc}}^{\mathrm{scalar}}\left[f^{(0)},f^{(1)}\right]+I_{\mathrm{PB}}^{\mathrm{scalar}}\left[f^{(0)}\right],\label{eq:next-to-leading-f}
\end{equation}
where the subscript 'qc' represents the quasi-classical contribution
and 'PB' represents the contribution from the Poisson bracket term.
The quasi-classical part can be obtained from Eq. (\ref{eq:leading-NJL-boltz-f})
by replacing all zeroth order distributions $\overline{f}_{i}^{(0)}$
and $\mathbf{S}_{i}^{(0)}$ by $\overline{f}_{i}=\overline{f}_{i}^{(0)}+\overline{f}_{i}^{(1)}$
and $\mathbf{S}_{i}=\mathbf{S}_{i}^{(0)}+\mathbf{S}_{i}^{(1)}$ respectively
and expanding it to the first order, the result is 
\begin{eqnarray}
I_{\mathrm{qc}}^{\mathrm{scalar}}\left[f^{(0)},f^{(1)}\right] & = & 4\hbar(g_{0}-3g_{\sigma})^{2}\int\frac{d^{3}\mathbf{p}_{1}}{(2\pi\hbar)^{3}}\frac{d^{3}\mathbf{p}_{2}}{(2\pi\hbar)^{3}}\frac{d^{3}\mathbf{p}_{3}}{(2\pi\hbar)^{3}}(2\pi\hbar)^{4}\delta^{(4)}(p+p_{3}-p_{2}-p_{1})\nonumber \\
 &  & \times\left\{ \overline{f}_{1}^{(1)}\left[\overline{f}_{2}^{(0)}+\overline{f}_{3}^{(0)}\overline{f}_{p}^{(0)}-\overline{f}_{2}^{(0)}\overline{f}_{3}^{(0)}-\overline{f}_{2}^{(0)}\overline{f}_{p}^{(0)}\right]\right.\nonumber \\
 &  & +\overline{f}_{2}^{(1)}\left[\overline{f}_{1}^{(0)}+\overline{f}_{3}^{(0)}\overline{f}_{p}^{(0)}-\overline{f}_{1}^{(0)}\overline{f}_{3}^{(0)}-\overline{f}_{1}^{(0)}\overline{f}_{p}^{(0)}\right]\nonumber \\
 &  & -\overline{f}_{3}^{(1)}\left[\overline{f}_{p}^{(0)}+\overline{f}_{1}^{(0)}\overline{f}_{2}^{(0)}-\overline{f}_{p}^{(0)}\overline{f}_{1}^{(0)}-\overline{f}_{p}^{(0)}\overline{f}_{2}^{(0)}\right]\nonumber \\
 &  & -\overline{f}_{p}^{(1)}\left[\overline{f}_{3}^{(0)}+\overline{f}_{1}^{(0)}\overline{f}_{2}^{(0)}-\overline{f}_{3}^{(0)}\overline{f}_{1}^{(0)}-\overline{f}_{3}^{(0)}\overline{f}_{2}^{(0)}\right]\nonumber \\
 &  & -(1-\overline{f}_{3}^{(0)}-\overline{f}_{p}^{(0)})\mathbf{S}_{1}^{(0)}\cdot\mathbf{S}_{2}^{(1)}-(1-\overline{f}_{3}^{(0)}-\overline{f}_{p}^{(0)})\mathbf{S}_{1}^{(1)}\cdot\mathbf{S}_{2}^{(0)}\nonumber \\
 &  & +(1-\overline{f}_{1}^{(0)}-\overline{f}_{2}^{(0)})\mathbf{S}_{3}^{(1)}\cdot\mathbf{S}_{p}^{(0)}+(1-\overline{f}_{1}^{(0)}-\overline{f}_{2}^{(0)})\mathbf{S}_{3}^{(0)}\cdot\mathbf{S}_{p}^{(1)}\nonumber \\
 &  & \left.+\mathbf{S}_{1}^{(0)}\cdot\mathbf{S}_{2}^{(0)}(\overline{f}_{3}^{(1)}+\overline{f}_{p}^{(1)})-(\overline{f}_{1}^{(1)}+\overline{f}_{2}^{(1)})\mathbf{S}_{3}^{(0)}\cdot\mathbf{S}_{p}^{(0)}\right\} .\label{eq:njl-scalar-qc}
\end{eqnarray}
We see in each term there is only one first order distribution with
all other distributions being of zeroth order. One can prove 
\begin{equation}
I_{\mathrm{PB}}^{\mathrm{scalar}}\left[f^{(0)}\right]=0,\label{eq:i-scalar-pb}
\end{equation}
for the scalar part of the Boltzmann equation.

In order to prove Eq. (\ref{eq:i-scalar-pb}), we use the following
property for the trace of Pauli matrices
\begin{eqnarray}
\mathrm{Tr}\left[\sigma_{i}\sigma_{j}\cdots\sigma_{k}\right] & = & (-1)^{n}\mathrm{Tr}\left[\sigma_{k}\cdots\sigma_{j}\sigma_{i}\right],
\end{eqnarray}
where the $n$ is number of Pauli matrices. To prove the above relation,
we insert $C^{2}=-1$ between Pauli matrices 
\begin{align}
 & (-1)^{n}\mathrm{Tr}\left[C\sigma_{i}CC\sigma_{j}C\cdots C\sigma_{k}C\right]\nonumber \\
 & =(-1)^{n}\mathrm{Tr}[\sigma_{i}^{T}\sigma_{j}^{T}\cdots\sigma_{k}^{T}]=(-1)^{n}\mathrm{Tr}\left[\sigma_{k}\cdots\sigma_{j}\sigma_{i}\right],\label{eq:pauli-rel}
\end{align}
where $C=i\sigma^{2}$ and $C\boldsymbol{\sigma}C=\boldsymbol{\sigma}^{T}$.


With Eq. (\ref{eq:pauli-rel}) we consider the quantities in the Poisson
bracket collision term $I_{\mathrm{PB}}^{\mathrm{scalar}}\left[f^{(0)}\right]$,
\begin{align}
I_{1} & =g_{c1}g_{c2}\mathrm{Im}\left[\mathrm{Tr}\left(\Gamma^{(c2)}G_{1}\Gamma^{(c1)}G_{3}\right)\mathrm{Tr}\left(\Gamma^{(c2)}G_{2}\Gamma^{(c1)}G_{p}\right)\right],\nonumber \\
I_{2} & =g_{c1}g_{c2}\mathrm{Im}\mathrm{Tr}\left(\Gamma^{(c2)}G_{1}\Gamma^{(c1)}G_{3}\Gamma^{(c2)}G_{2}\Gamma^{(c1)}G_{p}\right).
\end{align}
Note that $G_{i}$ $(i=1,2,3,p)$ can be either $G^{\lessgtr}(x,p_{i})$
or derivatives of $G^{\lessgtr}(x,p_{i})$. We know that $G_{i}$
contain the scalar and polarization parts, so we can express $G_{i}=\sum_{b_{i}=1,0}G_{i}(b_{i})$,
where $G_{i}(1)$ denotes its scalar part and $G_{i}(0)$ denotes
its polarization part. Let us work on $I_{1}$, 
\begin{eqnarray}
I_{1} & = & g_{c1}g_{c2}\mathrm{Im}\left[\mathrm{Tr}\left(\Gamma^{(c2)}G_{1}\Gamma^{(c1)}G_{3}\right)\mathrm{Tr}\left(\Gamma^{(c2)}G_{2}\Gamma^{(c1)}G_{p}\right)\right]\nonumber \\
 & = & g_{c1}g_{c2}\sum_{b_{1},b_{2},b_{3},b}\mathrm{Im}\left\{ \mathrm{Tr}\left[\Gamma^{(c2)}G_{1}(b_{1})\Gamma^{(c1)}G_{3}(b_{3})\right]\mathrm{Tr}\left[\Gamma^{(c2)}G_{2}(b_{2})\Gamma^{(c1)}G_{p}(b)\right]\right\} \nonumber \\
 & = & g_{c1}g_{c2}\sum_{n=\mathrm{odd}}\mathrm{Im}\left\{ \mathrm{Tr}\left[\Gamma^{(c2)}G_{1}(b_{1})\Gamma^{(c1)}G_{3}(b_{3})\right]\mathrm{Tr}\left[\Gamma^{(c2)}G_{2}(b_{2})\Gamma^{(c1)}G_{p}(b)\right]\right\} \nonumber \\
 & \downarrow & \mathrm{insert}\;C^{2}\nonumber \\
 & = & g_{c1}g_{c2}\sum_{n=\mathrm{odd}}(-1)^{n}\mathrm{Im}\left\{ \mathrm{Tr}\left[\Gamma^{(c1)}G_{1}(b_{1})\Gamma^{(c2)}G_{3}(b_{3})\right]\mathrm{Tr}\left[\Gamma^{(c1)}G_{2}(b_{2})\Gamma^{(c2)}G_{p}(b)\right]\right\} \nonumber \\
 & = & -g_{c1}g_{c2}\sum_{n=\mathrm{odd}}\mathrm{Im}\left\{ \mathrm{Tr}\left[\Gamma^{(c2)}G_{1}(b_{1})\Gamma^{(c1)}G_{3}(b_{3})\right]\mathrm{Tr}\left[\Gamma^{(c2)}G_{2}(b_{2})\Gamma^{(c1)}G_{p}(b)\right]\right\} ,\label{eq:i1-si}
\end{eqnarray}
where $n=b_{1}+b_{2}+b_{3}+b$ is the number of scalar parts in two
traces, and we have interchanged $c1\leftrightarrow c2$ in the last
equality since a summation over $c1$ and $c2$ is implied. Note that
even or odd $n$ also indicates even or odd number of Pauli matrices
in two traces respectively. Only when $n$ is odd does the product
of two traces have an imaginary part. By comparing the last equality
with the second one of Eq. (\ref{eq:i1-si}) we arrive at $I_{1}=0$.
For $I_{2}$, we have 
\begin{eqnarray}
I_{2} & = & g_{c1}g_{c2}\mathrm{Im}\mathrm{Tr}\left(\Gamma^{(c2)}G_{1}\Gamma^{(c1)}G_{3}\Gamma^{(c2)}G_{2}\Gamma^{(c1)}G_{p}\right)\nonumber \\
 & = & g_{c1}g_{c2}\sum_{b_{1},b_{2},b_{3},b}\mathrm{Im}\mathrm{Tr}\left[\Gamma^{(c2)}G_{1}(b_{1})\Gamma^{(c1)}G_{3}(b_{3})\Gamma^{(c2)}G_{2}(b_{2})\Gamma^{(c1)}G_{p}(b)\right]\nonumber \\
 & = & g_{c1}g_{c2}\sum_{n=\mathrm{odd}}\mathrm{Im}\mathrm{Tr}\left[\Gamma^{(c2)}G_{1}(b_{1})\Gamma^{(c1)}G_{3}(b_{3})\Gamma^{(c2)}G_{2}(b_{2})\Gamma^{(c1)}G_{p}(b)\right]\nonumber \\
 & \downarrow & \mathrm{insert}\;C^{2}\nonumber \\
 & = & g_{c1}g_{c2}\sum_{n=\mathrm{odd}}(-1)^{n}\mathrm{Im}\mathrm{Tr}\left[\Gamma^{(c1)}G_{2}(b_{2})\Gamma^{(c2)}G_{3}(b_{3})\Gamma^{(c1)}G_{1}(b_{1})\Gamma^{(c2)}G_{p}(b)\right]\nonumber \\
 & = & -g_{c1}g_{c2}\sum_{n=\mathrm{odd}}\mathrm{Im}\mathrm{Tr}\left[\Gamma^{(c2)}G_{1}(b_{1})\Gamma^{(c1)}G_{3}(b_{3})\Gamma^{(c2)}G_{2}(b_{2})\Gamma^{(c1)}G_{p}(b)\right],\label{eq:i2-si}
\end{eqnarray}
where $n=b_{1}+b_{2}+b_{3}+b$ is the number of scalar parts in the
trace, in the final equality we have interchanged $c1\leftrightarrow c2$
and $G_{1}(b_{1})\leftrightarrow G_{2}(b_{2})$ since a summation
over $c1$ and $c2$ is implied and there is a symmetry in the labels
1 and 2 (in the integration over $p_{1}$ and $p_{2}$ and the summation
over $b_{1}$ and $b_{2}$). The odd/even $n$ also corresponds to
odd/even number of Pauli matrices inside the trace. Only when $n$
is odd does the trace have an imaginary part. By comparing the last
equality with the second one of Eq. (\ref{eq:i2-si}), we obtain $I_{2}=0$.


We now derive the Boltzmann equation for the polarization distribution
at the next-to-leading order. The contributions can be grouped into
the local (quasi-classical) and nonlocal parts of the collision term.
The local part contains no space-time derivatives and can be obtained
from Eq. (\ref{eq:leading-NJL-pol-s}) by the replacement $\overline{f}_{i}^{(0)}\rightarrow\overline{f}_{i}=\overline{f}_{i}^{(0)}+\overline{f}_{i}^{(1)}$
and $\mathbf{S}_{i}^{(0)}\rightarrow\mathbf{S}_{i}=\mathbf{S}_{i}^{(0)}+\mathbf{S}_{i}^{(1)}$
in the collision term and then by expanding the collision term to
the next-to-leading order. The nonlocal part comes from the Poisson
bracket term with space-time derivatives. Special care should be taken
for the derivatives in $\partial_{p_{0}}$ and $\nabla_{p}$ which
act on the two-point function $G^{\lessgtr}(x,p)$, giving terms with
$\delta^{\prime}(p_{0}-\mathbf{p}^{2}/2m)$. These terms belong to
off-shell contributions which we will neglect in this section and
leave to the next section for treatment. Substituting Eq. (\ref{eq:on-shell-g})
into Eq. (\ref{eq:boltzman-s}), we obtain the on-shell Boltzmann
equation for the polarization distribution at the next-to-leading
order
\begin{eqnarray}
\hbar\left(\partial_{t}+\frac{1}{m}\mathbf{p}\cdot\nabla_{x}\right)\mathbf{S}_{p}^{a,(1)} & = & I_{\mathrm{qc}}^{\mathrm{pol}}\left[f^{(0)},f^{(1)}\right]+I_{\mathrm{PB}}^{\mathrm{pol}}\left[f^{(0)}\right].\label{eq:NL-NJL-bol-s-1}
\end{eqnarray}
The explicit form of the local part (quasi-classical) of the collision
term is given by
\begin{eqnarray}
I_{\mathrm{qc}}^{\mathrm{pol}}\left[f^{(0)},f^{(1)}\right] & = & 4\hbar(g_{0}-3g_{\sigma})^{2}\int\frac{d^{3}\mathbf{p}_{1}}{(2\pi\hbar)^{3}}\frac{d^{3}\mathbf{p}_{2}}{(2\pi\hbar)^{3}}\frac{d^{3}\mathbf{p}_{3}}{(2\pi\hbar)^{3}}(2\pi\hbar)^{4}\delta^{(4)}(p+p_{3}-p_{2}-p_{1})\nonumber \\
 &  & \times\left\{ \mathbf{S}_{3}^{a,(0)}\left[\overline{f}_{p}^{(1)}(1-\overline{f}_{1}^{(0)}-\overline{f}_{2}^{(0)})+\overline{f}_{1}^{(1)}(\overline{f}_{2}^{(0)}-\overline{f}_{p}^{(0)})+\overline{f}_{2}^{(1)}(\overline{f}_{1}^{(0)}-\overline{f}_{p}^{(0)})\right]\right.\nonumber \\
 &  & -\mathbf{S}_{p}^{a,(0)}\left[\overline{f}_{3}^{(1)}(1-\overline{f}_{1}^{(0)}-\overline{f}_{2}^{(0)})+\overline{f}_{1}^{(1)}(\overline{f}_{2}^{(0)}-\overline{f}_{3}^{(0)})+\overline{f}_{2}^{(1)}(\overline{f}_{1}^{(0)}-\overline{f}_{3}^{(0)})\right]\nonumber \\
 &  & +\mathbf{S}_{3}^{a,(1)}\left(\overline{f}_{p}^{(0)}+\overline{f}_{1}^{(0)}\overline{f}_{2}^{(0)}-\overline{f}_{p}^{(0)}\overline{f}_{1}^{(0)}-\overline{f}_{p}^{(0)}\overline{f}_{2}^{(0)}\right)\nonumber \\
 &  & -\mathbf{S}_{p}^{a,(1)}\left(\overline{f}_{3}^{(0)}+\overline{f}_{1}^{(0)}\overline{f}_{2}^{(0)}-\overline{f}_{3}^{(0)}\overline{f}_{1}^{(0)}-\overline{f}_{3}^{(0)}\overline{f}_{2}^{(0)}\right)\nonumber \\
 &  & +\left(\mathbf{S}_{1}^{(1)}\cdot\mathbf{S}_{2}^{(0)}+\mathbf{S}_{1}^{(0)}\cdot\mathbf{S}_{2}^{(1)}\right)\left(\mathbf{S}_{p}^{a,(0)}-\mathbf{S}_{3}^{a,(0)}\right)\nonumber \\
 &  & \left.+\mathbf{S}_{1}^{(0)}\cdot\mathbf{S}_{2}^{(0)}\left(\mathbf{S}_{p}^{a,(1)}-\mathbf{S}_{3}^{a,(1)}\right)\right\} .\label{eq:NL-NJL-pol-qc}
\end{eqnarray}
One can check that $I_{\mathrm{qc}}^{\mathrm{pol}}\left[f^{(0)},f^{(1)}\right]$
is local and on-shell.


The non-local part of the collision term contains on-shell and off-shell
contributions, as we have mentioned, in this section we focus on the
on-shell contribution and will treat the off-shell one in Sect. \ref{sec:off-shell-part}.
The explicit form of the on-shell Poisson bracket term reads 
\begin{eqnarray}
I_{\mathrm{PB}}^{\mathrm{pol}}\left[f^{(0)}\right] & \approx & 2\hbar^{2}\left(g_{0}-3g_{\sigma}\right)^{2}\epsilon_{aij}\int\frac{d^{3}p_{1}}{(2\pi\hbar)^{3}}\frac{d^{3}p_{2}}{(2\pi\hbar)^{3}}\frac{d^{3}p_{3}}{(2\pi\hbar)^{3}}(2\pi\hbar)^{4}\delta^{(4)}(p+p_{3}-p_{1}-p_{2})\nonumber \\
 &  & \times\left\{ \mathbf{S}_{3}^{(0),i}\left[\nabla_{x}\left(\overline{f}_{1}^{(0)}+\overline{f}_{2}^{(0)}\right)\cdot\nabla_{p}\mathbf{S}_{p}^{(0),j}-\left(\nabla_{p_{1}}\overline{f}_{1}^{(0)}+\nabla_{p_{2}}\overline{f}_{2}^{(0)}\right)\cdot\nabla_{x}\mathbf{S}_{p}^{(0),j}\right]\right.\nonumber \\
 &  & \left.+\left(1-\overline{f}_{1}^{(0)}-\overline{f}_{2}^{(0)}\right)\left(\nabla_{p_{3}}\mathbf{S}_{3}^{(0),i}\cdot\nabla_{x}\mathbf{S}_{p}^{(0),j}-\nabla_{x}\mathbf{S}_{3}^{(0),i}\cdot\nabla_{p}\mathbf{S}_{p}^{(0),j}\right)\right\} .\label{eq:nl-njl-pol-on}
\end{eqnarray}
Here we have neglected a term with 
\[
\partial_{p_{0}}\delta^{(4)}(p+p_{3}-p_{1}-p_{2})\epsilon_{aij}\mathbf{S}_{3}^{(0),j}\left(\partial_{t}+\frac{\mathbf{p}}{m}\cdot\nabla_{x}\right)\mathbf{S}_{p}^{j,(0)}
\]
in the integrand which is of order $\mathcal{O}(\hbar^{2}g_{\mathrm{couple}}^{4})$
by Eq. (\ref{eq:leading-NJL-pol-s}), while the terms in Eq. (\ref{eq:nl-njl-pol-on})
are all of order $\mathcal{O}(\hbar^{2}g_{\mathrm{couple}}^{2})$,
where $g_{\mathrm{couple}}=g_{0}$ or $g_{\sigma}$ denotes the coupling
constant. The collision term in Eq. (\ref{eq:nl-njl-pol-on}) is non-local
for it contains derivatives of space-time. From Eq. (\ref{eq:nl-njl-pol-on}),
we see that the collision term vanishes if there is no polarization
density at the leading order, i.e., $\mathbf{S}_{p}^{a,(0)}=0$. This
is the result of the non-relativistic coupling of the NJL type. In
contrast it has been proved in Ref. \citep{Sheng:2021kfc} that the
polarization can be generated from the Poisson bracket term even without
polarization density at the leading order in a relativistic NJL model.


\section{Spin boltzmann equations: off-shell parts}

\label{sec:off-shell-part}In this section we will investigate off-shell
contributions. From Eq. (\ref{eq:on-shell}) and (\ref{eq:kb-evo})
$G^{<}(x,p)$ actually contains an off-shell part $G_{\mathrm{off}}^{<}(x,p)$
besides the on-shell part in (\ref{eq:on-shell-g}) which we denote
as $G_{\mathrm{on}}^{<}(x,p)$, so do $I_{\mathrm{coll}}$ and $I_{\mathrm{coll}}^{\dagger}$.

Now we try to derive the connection between $G_{\mathrm{on}}^{<}(x,p)$
and $G_{\mathrm{off}}^{<}(x,p)$. From Eq. (\ref{eq:on-shell}) we
obtain
\begin{equation}
G_{\mathrm{off}}^{<}(x,p)=\frac{1}{2p_{0}-\mathbf{p}^{2}/m}\left(I_{\mathrm{coll}}+I_{\mathrm{coll}}^{\dagger}\right)+\mathcal{O}(\hbar^{2}).\label{eq:off-shell-3}
\end{equation}
Note that the on-shell part $I_{\mathrm{coll}}^{\mathrm{on}}$ contains
$\delta\left[p_{0}-\mathbf{p}^{2}/(2m)\right]$ which, combining $\left[p_{0}-\mathbf{p}^{2}/(2m)\right]^{-1}$,
gives the derivative of the delta-function, $\delta^{\prime}\left[p_{0}-\mathbf{p}^{2}/(2m)\right]$.
The explicit form of $G_{\mathrm{off}}^{<}(x,p)$ can be determined
by the collision term from Eq. (\ref{eq:off-shell-3}) and is at least
of the same order as $\hbar G_{\mathrm{on}}^{\lessgtr(1)}$ in Eq.
(\ref{eq:g1}). We can express $G_{\mathrm{off}}^{<}(x,p)$ as 
\begin{eqnarray}
G_{\mathrm{off}}^{<}(x,p) & = & G_{\mathrm{off}}^{>}(x,p)\nonumber \\
 & = & -\hbar(2\pi\hbar)\partial_{p_{0}}\delta\left(p_{0}-\frac{\mathbf{p}^{2}}{2m}\right)\sum_{s_{1},s_{2}=\pm}\chi(s_{1})\chi^{\dagger}(s_{2})f_{s_{1}s_{2}}^{\mathrm{off}}(x,\mathbf{p})\nonumber \\
 & = & -\hbar(2\pi\hbar)\partial_{p_{0}}\delta\left(p_{0}-\frac{\mathbf{p}^{2}}{2m}\right)\left[\overline{f}_{\mathrm{off}}(x,\mathbf{p})+\boldsymbol{\sigma}\cdot\mathbf{S}_{\mathrm{off}}(x,\mathbf{p})\right]\nonumber \\
 & = & -\hbar(2\pi\hbar)\partial_{p_{0}}\delta\left(p_{0}-\frac{\mathbf{p}^{2}}{2m}\right)\boldsymbol{\sigma}\cdot\mathbf{S}_{\mathrm{off}}(x,\mathbf{p}),\label{eq:g-off-f}
\end{eqnarray}
where we used $\overline{f}_{\mathrm{off}}(x,\mathbf{p})=0$ and 
\begin{eqnarray}
\mathbf{S}_{\mathrm{off}}^{a}(x,\mathbf{p}) & = & 2(g_{0}-3g_{\sigma})^{2}\epsilon^{aij}\int\frac{d^{4}p_{1}}{(2\pi\hbar)^{3}}\frac{d^{4}p_{2}}{(2\pi\hbar)^{3}}\frac{d^{4}p_{3}}{(2\pi\hbar)^{3}}(2\pi\hbar)^{4}\delta(p+p_{3}-p_{2}-p_{1})\nonumber \\
 &  & \times\delta\left(p_{1}^{0}-\frac{\mathbf{p}_{1}^{2}}{2m}\right)\delta\left(p_{2}^{0}-\frac{\mathbf{p}_{2}^{2}}{2m}\right)\delta\left(p_{3}^{0}-\frac{\mathbf{p}_{3}^{2}}{2m}\right)\left(1-\overline{f}_{1}^{(0)}-\overline{f}_{2}^{(0)}\right)\mathbf{S}_{3}^{i,(0)}\mathbf{S}_{p}^{j,(0)},\label{eq:off-shell-2}
\end{eqnarray}
following Eq. (\ref{eq:off-shell-3}) at the leading order. Actually
$G_{\mathrm{off}}^{<}$ in the form of (\ref{eq:g-off-f}) would be
added to the left-hand-side of Eq. (\ref{eq:boltzman-s}) at the next-to-leading
order {[}since $\overline{f}_{\mathrm{off}}(x,\mathbf{p})=0$, there
is no off-shell correction to Eq. (\ref{eq:boltzman-f}){]}, while
there is also the off-shell part from the Poisson bracket term in
the right-hand-side. It can be shown that the off-shell terms in both
sides of Eq. (\ref{eq:boltzman-s}) are equal at $\mathcal{O}(\hbar^{2}g_{\mathrm{couple}}^{2})$
and thus drop out from the equation to leave Eq. (\ref{eq:NL-NJL-bol-s-1})
for the on-shell part.


From Eq. (\ref{eq:kb-evo}), we obtain 
\begin{eqnarray}
\hbar\left(\partial_{t}+\frac{1}{m}\mathbf{p}\cdot\nabla_{x}\right)G_{\mathrm{on}}^{<}(x,p) & = & -i\left(I_{\mathrm{coll}}-I_{\mathrm{coll}}^{\dagger}\right)-\hbar\left(\partial_{t}+\frac{\mathbf{p}}{m}\cdot\nabla_{x}\right)G_{\mathrm{off}}^{<}(x,p)\nonumber \\
 & = & -i\left(I_{\mathrm{coll}}^{\mathrm{on}}-I_{\mathrm{coll}}^{\mathrm{on}\dagger}\right)-i\left(I_{\mathrm{coll}}^{\mathrm{off}}-I_{\mathrm{coll}}^{\mathrm{off}\dagger}\right)\nonumber \\
 &  & -\hbar\frac{1}{2p_{0}-\mathbf{p}^{2}/m}\left(\partial_{t}+\frac{\mathbf{p}}{m}\cdot\nabla_{x}\right)\left(I_{\mathrm{coll}}+I_{\mathrm{coll}}^{\dagger}\right)+\mathcal{O}(\hbar^{3}).\label{eq:on-off-rel}
\end{eqnarray}
Now we act the operator in the left-hand side of Eq. (\ref{eq:kb-g-less-g-larger})
to the right-hand-side of Eq. (\ref{eq:kb-g-left}) and obtain 
\begin{equation}
\left(-i\hbar\frac{\partial}{\partial t_{2}}+\frac{\hbar^{2}}{2m}\nabla_{x2}^{2}\right)I_{\mathrm{coll}}(x_{1},x_{2})=\mathcal{O}(\hbar^{2}g_{\mathrm{couple}}^{4}).\label{eq:on-off}
\end{equation}
Note that the operator only acts on Green functions in $I_{\mathrm{coll}}$
(\ref{eq:i-coll}) instead of self-energies since only Green functions
depend on $x_{2}$, which gives a contribution of $\mathcal{O}(\hbar^{2}g_{\mathrm{couple}}^{4})$
using Eq. (\ref{eq:kb-g-less-g-larger}). Taking a Wigner transform
of Eq. (\ref{eq:on-off}) we obtain an equation for $I_{\mathrm{coll}}^{\mathrm{on}}$
and $I_{\mathrm{coll}}^{\mathrm{off}}$ 
\begin{eqnarray}
I_{\mathrm{coll}}^{\mathrm{off}} & = & \frac{1}{2p_{0}-\mathbf{p}^{2}/m}i\hbar\left(\partial_{t}+\frac{\mathbf{p}}{m}\cdot\nabla_{x}\right)I_{\mathrm{coll}}+\mathcal{O}(\hbar^{2}g_{\mathrm{couple}}^{4})+\mathcal{O}(\hbar^{3}g_{\mathrm{couple}}^{2}),\label{eq:i-off}
\end{eqnarray}
where the first term is at least of $\mathcal{O}(\hbar^{2}g_{\mathrm{couple}}^{2})$.
Taking a Hermitian conjugate of the above equation leads to
\begin{eqnarray}
I_{\mathrm{coll}}^{\mathrm{off}\dagger} & = & -\frac{1}{2p_{0}-\mathbf{p}^{2}/m}i\hbar\left(\partial_{t}+\frac{\mathbf{p}}{m}\cdot\nabla_{x}\right)I_{\mathrm{coll}}^{\dagger}+\mathcal{O}(\hbar^{2}g_{\mathrm{couple}}^{4})+\mathcal{O}(\hbar^{3}g_{\mathrm{couple}}^{2}).\label{eq:i-off-1}
\end{eqnarray}
Substituting Eqs. (\ref{eq:i-off}) and (\ref{eq:i-off-1}) into (\ref{eq:on-off-rel})
we arrive at an equation for the on-shell part of the two-point function
\begin{equation}
\hbar\left(\partial_{t}+\frac{\mathbf{p}}{m}\cdot\nabla_{x}\right)G_{\mathrm{on}}^{<}(x,p)=-i\left(I_{\mathrm{coll}}^{\mathrm{on}}-I_{\mathrm{coll}}^{\mathrm{on}\dagger}\right)+\mathcal{O}(\hbar^{2}g_{\mathrm{couple}}^{4})+\mathcal{O}(\hbar^{3}g_{\mathrm{couple}}^{2}).\label{eq:decom-on-off}
\end{equation}
The above equation for on-shell parts is the foundation for discussions
in Sect. \ref{sec:NJL-on-shell}, in which the index ``on'' has
been suppressed for notational simplicity. In this section we have
resumed the use of index ``on'' to denote $G_{\mathrm{on}}^{\lessgtr}(x,p)$
and $I_{\mathrm{coll}}^{\mathrm{on}}$.


Let us analyze $I_{\mathrm{coll}}^{\mathrm{off}}$ in Eq. (\ref{eq:i-off}).
From Eq. (\ref{eq:coll-on-shell}), $I_{\mathrm{coll}}$ has two parts:
one is the local term or quasi-classical term $I_{\mathrm{qc}}$ which
is at least of $\mathcal{O}(\hbar)$, and the other is the nonlocal
term with Poisson brackets $I_{\mathrm{PB}}$ which is at least of
$\mathcal{O}(\hbar^{2})$. The leading contribution of $I_{\mathrm{coll}}^{\mathrm{off}}$
is of $\mathcal{O}(\hbar^{2}g_{\mathrm{couple}}^{2})$ from $I_{\mathrm{PB}}$
containing derivatives $\partial_{p_{0}}$ and $\nabla_{p}$ acting
on $G^{\lessgtr}(x,p)$, while the contribution from $G_{\mathrm{off}}^{<}$
in (\ref{eq:g-off-f}) to $I_{\mathrm{coll}}^{\mathrm{off}}$ through
the expansion of $I_{\mathrm{qc}}$ is of $\mathcal{O}(\hbar^{2}g_{\mathrm{couple}}^{4})$
which is in higher order. Note that the leading contribution in $I_{\mathrm{coll}}^{\mathrm{off}}$
from $I_{\mathrm{PB}}$ is in the same order as the first term in
the right-hand-side of Eq. (\ref{eq:i-off}). So at $\mathcal{O}(\hbar^{2}g_{\mathrm{couple}}^{2})$,
Eq. (\ref{eq:i-off}) gives a constraint for MVSD $f^{(0)}$ or equivalently
$\overline{f}^{(0)}$ and $\mathbf{S}^{(0)}$ in addition to Eqs.
(\ref{eq:leading-NJL-boltz-f}) and (\ref{eq:leading-NJL-pol-s}).


\section{Comments on nuclear force through OBEP}

In order to apply our theory to a non-relativistic nucleon system
in low energy collisions, one has to go beyond the contact interaction
of the NJL type and consider nuclear force as interaction. The main
features of nuclear force can be effectively decribed by one boson
exchange potential (OBEP) \citep{Bryan:1969mp,Machleidt:1989tm}.
The OBEPs through scalar, pseudoscalar, and vector meson exchanges
have terms with operators $\boldsymbol{\sigma}_{1}\cdot\boldsymbol{\sigma}_{2}$,
$\mathbf{L}\cdot\mathbf{S}$ and $S_{12}$ defined as $\mathbf{L}\equiv-i\mathbf{r}\times\nabla$,
$\mathbf{S}\equiv(\boldsymbol{\sigma}_{1}+\boldsymbol{\sigma}_{2})/2$,
and 
\begin{equation}
S_{12}\equiv\frac{1}{r^{2}}\left[3(\boldsymbol{\sigma}_{1}\cdot\mathbf{r})(\boldsymbol{\sigma}_{2}\cdot\mathbf{r})-r^{2}\boldsymbol{\sigma}_{1}\cdot\boldsymbol{\sigma}_{2}\right],
\end{equation}
where $\mathbf{r}\equiv\mathbf{x}-\mathbf{x}^{\prime}$. So OBEPs
are nolocal and contain couplings between spin and coordinate (equivalently
spin and momentum). In this case the spin is not a conserved quantity
as in the NJL-like model. It may be converted from local orbital angular
momentum or local vorticity \citep{Weickgenannt:2020aaf,Weickgenannt:2021cuo}.
The extension to the OBEP is much more complicated and beyond the
scope of this work. It will be reserved for a future study. 


\section{Summary}

\label{sec:Summary}We derive spin Boltzmann equations for non-relativistic
spin-1/2 fermions from the KB equation in the CTP formalism. The non-relativistic
model is similar to the NJL model with four-fermion contact interaction
which conserves spins in particle scatterings. The great merit of
the model is that the spin matrix element in the collision term can
be completely worked out and be put into a compact form. One can clearly
see how spins are coupled in two-to-two scatterings of particles.
In contrast it is hard to envisage the structure of the spin matrix
element which is much more complicated in the relativistic theory
\citep{Sheng:2021kfc}.

Starting from the non-relativistic Lagrangian, the KB equation is
derived from the Dyson-Schwinger equation defined on the CTP. The
spin Boltzmann equations for the particle number and spin distribution
are derived based on Wigner functions and the KB equation. Since the
spin polarization is a quantum effect, we make an expansion in the
Planck constant $\hbar$ for all quantities in the spin Boltzmann
equation. At the leading order, the equilibrium spin distribution
can be obtained under the condition of the vanishing collision term
for the spin phase space density. A spin chemical potential emerges
in the equilibrium spin distribution which is a natural consequence
of spin conservation. The off-shell parts of spin Boltzmann equations
are also discussed. The work can be extended to a system of nucleons
which interact via nuclear forces in low energy heavy-ion collisions.


\textit{Acknowledgement}. This work is supported in part by the Strategic
Priority Research Program of the Chinese Academy of Sciences (CAS)
under Grant No. XDB34030102, and by the National Natural Science Foundation
of China (NSFC) under Grants No. 12135011, 11890713 (a subgrant of
11890710).

\bibliographystyle{h-physrev}
\bibliography{citation}

\end{document}